\documentclass[aps,prl,twocolumn,showpacs,preprintnumbers,amsmath,amssymb,floatfix,nofootinbib]{revtex4-1}
\usepackage{graphicx}

\usepackage{color}
\usepackage{colordvi}
\usepackage[normalem]{ulem}
\usepackage{cancel}
\newcommand{\beqn}{\begin{eqnarray}}
\newcommand{\eeqn}{\end{eqnarray}}
\newcommand{\be}{\begin{equation}}
\newcommand{\ee}{\end{equation}}

\newcommand{\ba}{\begin{array}{c}}
\newcommand{\bat}{\begin{array}{cc}}
\newcommand{\ea}{\end{array}}

\newcommand{\bi}{\begin{itemize}}
\newcommand{\ei}{\end{itemize}}

\newcommand{\comment}[1]{}

\usepackage[utf8]{inputenc}
\usepackage{xcolor}
\usepackage{slashed}
\usepackage{graphicx}
\usepackage{amsmath}
\usepackage{setspace}

\begin{document}

\title{Radiative corrections to $\tau \to \pi (K) \nu_\tau [\gamma]$: a reliable new physics test }

\author{M. A. Arroyo-Ure\~na${}^{1}$}
%\email{}
\author{G.~%erardo
Hern\' andez-Tom\'e${}^{1}$}
%\email{}
\author{G.~%abriel
L\'opez-Castro${}^{1}$}
%\email{}
\author{P.~%ablo 
Roig${}^{1}$}
%\email{}
\author{I.~%gnasi
Rosell${}^{2}$}
%\email{}

\affiliation{${}^1$ Departamento de F\'\i sica, Centro de Investigaci\' on y de Estudios Avanzados del Instituto Polit\' ecnico Nacional, Apartado Postal 14-740, 07000 Ciudad de M\'exico, M\' exico }
\affiliation{${}^2$  Departamento de Matem\'aticas, F\'\i sica y Ciencias Tecnol\' ogicas, Universidad Cardenal Herrera-CEU, CEU Universities, 46115 Alfara del Patriarca, Val\`encia, Spain}

\begin{abstract}
The ratios  $R_{\tau/P}\equiv \Gamma(\tau \to P \nu_\tau [\gamma]) / \Gamma(P \to\mu \nu_\mu[\gamma])$ ($P=\pi, K$) provide sensitive tests of lepton universality  $\left|g_\tau/g_\mu\right|=1$ and are a useful tool for new physics searches. The radiative corrections to $R_{\tau/P}$ are computed following a large-$N_C$ expansion to deal with hadronic effects: Chiral Perturbation Theory is enlarged by including the lightest multiplets of spin-one heavy states such that the relevant Green functions are well-behaved at high energies. We find  $\delta R_{\tau/\pi}=(0.18\pm 0.57 )\%$  and $\delta R_{\tau/K}=(0.97\pm 0.58 )\%$, which imply $\left|g_\tau/g_\mu\right|_\pi=0.9964\pm 0.0038$ and $\left|g_\tau/g_\mu\right|_K=0.9857\pm 0.0078$, compatible with and at $1.8\sigma$ of lepton universality, respectively. We test unitarity and bind  non-standard effective interactions with the $\tau \to P \nu_\tau [\gamma]$ decays.
\end{abstract}

\maketitle

\vspace{-5cm}

{\it Introduction.} -- Lepton universality (LU) is a basic tenet of the Standard Model of particles interactions. A large diversity of weak interaction processes are compatible with the fact that lepton doublets have identical couplings $g_{\ell}$ to the $W$-boson. A few anomalies observed in semileptonic $B$ meson decays \cite{Albrecht:2021tul} seem to challenge this principle or require new non-universal weak interactions. Lower energy observables where very precise comparison of theory and experiments can be done, currently provide the most precise test of LU~\cite{Bryman:2021teu}.

In this work, we aim to test muon-tau lepton universality  through the ratio ($P=\pi,K$)~\cite{Marciano:1993sh, DF}
\begin{equation}\label{eq:MainDef}
 R_{\tau/P}\equiv \frac{\Gamma(\tau \to P \nu_\tau  [\gamma])}{\Gamma(P \to\mu \nu_\mu[\gamma])}=\bigg|\frac{g_\tau}{g_\mu}\Bigg|^2_P R_{\tau/P}^{(0)} \left(1+\delta R_{\tau/P}\right)\,,   
\end{equation}
where $g_\mu=g_\tau$ according to LU\footnote{The strong helicity suppression of the $P_{e2}$ decays disfavors verifying $g_e=g_\tau$ similarly.}, the radiative corrections are encoded in $\delta R_{\tau/P}$ and $R_{\tau/P}^{(0)}$ is the leading-order result,
\begin{equation}\label{eq:LO}
R_{\tau/P}^{(0)} = \frac{1}{2} \frac{M_\tau^3}{m_\mu^2m_P}\frac{(1-m_P^2/M_\tau^2)^2}{(1-m_\mu^2/m_P^2)^2}\, ,
\end{equation}
which is free from hadronic couplings and quark mixing angles.

$\delta R_{\tau/P}$ was calculated in Refs.~\cite{DF}, where the values $\delta R_{\tau/\pi}=(0.16\pm0.14)\%$ and $\delta R_{\tau/K}=(0.90\pm0.22)\%$ were reported. There are important reasons to address this analysis again:
\begin{enumerate}
\item {\bf Phenomenology}. Using $\delta R_{\tau/P}$ from \cite{DF}, the last HFLAV analysis \cite{Amhis:2019ckw} quoted $\left|g_\tau/g_\mu\right|_\pi=0.9958\pm0.0026$ and $\left|g_\tau/g_\mu\right|_K=0.9879\pm0.0063$, at $1.6\sigma$ and $1.9\sigma$ of LU ($1.4\sigma$ and $2.0\sigma$ in \cite{Pich:2013lsa},  making use of the PDG input \cite{PDG}). However, determinations of $\left|g_\tau/g_\mu\right|$ considering other observables are compatible with LU: the pure leptonic extraction via $\Gamma(\tau \to e \bar{\nu}_e \nu_\tau[\gamma])/\Gamma(\mu \to e \bar{\nu}_e \nu_\mu[\gamma])$, $\left|g_\tau/g_\mu\right|=1.0010\pm0.0014$ \cite{Amhis:2019ckw} ($1.0011\pm 0.0015$ in \cite{Pich:2013lsa}), and the weighted average of the recent $W$-boson decay determinations via $\Gamma(W \to\tau \nu_\tau)/\Gamma(W \to\mu \nu_\mu)$, $\left|g_\tau/g_\mu\right|=0.995\pm 0.006$~\cite{Aad:2020ayz, CMS:2021qxj}, agree remarkably with LU. Therefore, a closer look to the radiative corrections $\delta R_{\tau/P}$ is convenient to disentangle this disagreement.
\item {\bf Theory}. First of all, the hadronic form factors modeled in Ref.~\cite{DF} are different for real- and virtual-photon corrections. Furthermore, they do not satisfy the correct QCD short-distance behavior, violate unitarity,  analyticity and the chiral limit at leading non-trivial orders, and use a cutoff to regulate the loop integrals, separating unphysically long- and short-distance corrections. Finally, the uncertainties quoted in Ref.~\cite{DF} are unrealistic, since they are of the order of a purely Chiral Perturbation Theory result, that is, a computation which cannot include the $\tau$. Thus, a new analysis of $\delta R_{\tau/P}$ overcoming these problems is pressing from a theoretical point of view.
\end{enumerate}
Moreover, and as a by-product, an updated analysis of $\delta R_{\tau/P}$ would be useful to revisit the CKM unitarity test via $\left|V_{us}\right|$ in $\Gamma(\tau \to K \nu_\tau[\gamma])/\Gamma(\tau \to \pi \nu_\tau[\gamma])$ or $\Gamma(\tau \to K \nu_\tau[\gamma])$~\cite{Pich:2013lsa} and  improve the constraints on possible non-standard interactions affecting this ratio~\cite{EFTpions, EFTtaudecays}.\\ 

{\it $P \to\mu \nu_\mu[\gamma]$.} -- The inclusive $P_{\mu 2[\gamma]}$ decay rate can be analyzed unambiguously within the Standard Model (Chiral Perturbation Theory), being the estimation of the local counterterms the only model dependence. We follow the notation proposed in Ref.~\cite{Marciano:1993sh} and the numbers reported in Refs.~\cite{CR1,CR2}:
\begin{align}
&\Gamma_{P_{\mu 2 [\gamma]}} = 
\Gamma^{(0)}_{P_{\mu 2 }} \!
\Bigg\{ 1 + \frac{2 \, \alpha}{\pi}  \log \frac{m_Z}{m_\rho} \Bigg\}\!
\Bigg\{ 1 + \frac{\alpha}{\pi}  \,  F( m_\mu^2/m_P^2)  \Bigg\}
\nonumber \\ 
&\Bigg\{  1 - \frac{\alpha}{\pi}   \Bigg[ \frac{3}{2}  \log \frac{m_\rho}{m_P}   +   c_1^{(P)}  + \frac{m_\mu^2}{m_\rho^2}   \bigg(c_2^{(P)}  \, \log \frac{m_\rho^2}{m_\mu^2}   +  c_3^{(P)} \nonumber \\ &
+ c_4^{(P)} (m_\mu/m_P) \bigg)  -  \frac{m_P^2}{m_\rho^2} \,  \tilde{c}_{2}^{(P)}  \, \log \frac{m_\rho^2}{m_\mu^2}  \Bigg] \Bigg\} \,,
\label{eq:indrate}
\end{align}
where the first bracketed term is the universal short-distance electroweak correction (which cancels in the ratio  $R_{\tau/P}$), the second bracketed term is the universal long-distance correction (point-like approximation, originally calculated in Ref.~\cite{Kinoshita:1959ha} and to be given later), the third bracketed term includes the structure dependent contributions and $\Gamma^{(0)}_{P_{\mu 2 }}$ is the rate in absence of radiative corrections ($F_\pi\sim92$ MeV),
\begin{equation}
\Gamma^{(0)}_{P_{\mu 2}} \,=\,  \frac{G_F^2 |V_{uD}|^2  F_P^2 }{4 \pi} \, 
m_P  \, m_\mu^2  \, \left(1 - \frac{m_\mu^2}{m_P^2} \right)^2 \,, 
\end{equation}
being $D=d,s$ for $P=\pi,K$, respectively. The numerical values for $c_n^{(P)}$ are reported in Table~\ref{tab:tab1}~\cite{CR1,CR2}. Note that the most important uncertainties come from the estimations of the local counterterms, which were computed with a large-$N_C$ expansion of QCD where Chiral Perturbation Theory is enlarged by including the lightest multiplets of spin-one heavy states such that the relevant Green functions are well-behaved at high energies~\cite{RChT}.\\
\begin{table}[t!]
\begin{center}
\begin{tabular}{|c|c|c|}
\hline
  & $(P=\pi)$  & $(P=K)$    \\[5pt]
\hline 
$c_{1}^{(P)}$ & $-2.56\pm0.5_{\rm m }$ & $-1.98\pm0.5_{\rm m }$ \\
 $c_2^{(P)}$  & $5.2 \pm 0.4_{L_9} \pm 0.01_\gamma$  &  $4.3 \pm 0.4_{L_9} \pm 0.01_\gamma $ \\
 $c_3^{(P)}$   
&   
$ -10.5 \pm 2.3_{\rm m } \pm 0.53_{L_9} 
$
& 
$ -4.73 \pm 2.3_{ \rm m} \pm 0.28_{L_9}
$ 
\\
 $c_4^{(P)} $  &
$1.69 \pm 0.07_{L_9} $
&  
$ 0.22 \pm 0.01_{L_9} $ 
\\
$\tilde{c}_2^{(P)}$  &   0  &  $ (7.84 \pm 0.07_\gamma) \cdot 10^{-2}  $ \\
\hline
\end{tabular}
\end{center}
\caption{Numerical values for $c_n^{(P)}$ of (\ref{eq:indrate})~\cite{CR1,CR2} ($c_{1}^{(P)}$ from \cite{DescotesGenon:2005pw}). The uncertainties correspond to the input values  $L_9^r (\mu=m_\rho) = (6.9 \pm 0.7) \cdot 10^{-3} $, $\gamma\equiv F_{A}^\pi(0,0)/F_{V}^\pi(0,0)= 0.465 \pm 0.005$, and to the estimation of the counterterms (${\rm m}$, from matching), affecting only $c_1^{(P)}$ and $c_3^{(P)}$.}
\label{tab:tab1}
\end{table}

{\it $\tau \to P \nu_\tau  [\gamma]$.} -- $\tau$ decays must be scrutinized by using an effective approach encoding the hadronization of the QCD currents and we consider here the same large-$N_C$ expansion of QCD used in Refs.~\cite{CR1,CR2} to estimate the counterterms of $P_{\mu 2[\gamma]}$, quoted previously~\cite{RChT}. 
Similarly to (\ref{eq:indrate}), the decay rate can be organized as
\begin{align}
&\Gamma_{\tau_{P2[\gamma]}} = 
\Gamma^{(0)}_{\tau_{P2}}
\Bigg\{ 1 + \frac{2 \, \alpha}{\pi}  \log \frac{m_Z}{m_\rho} \Bigg\}\!
\Bigg\{ 1 + \frac{\alpha}{\pi}  \,  G (m_P^2/M_\tau^2)  \Bigg\}
\nonumber \\ 
&\qquad \qquad \Bigg\{  1 - \frac{3\alpha}{2\pi}  \log \frac{m_\rho}{M_\tau} +  \delta_{\tau P}\big|_{\mathrm{rSD}} +  \delta_{\tau P}\big|_{\mathrm{vSD}}\Bigg\} \,,
\label{eq:indratetau}
\end{align}
where again the point-like long-distance correction will be reported later, the structure dependent contributions have been split into the real-photon (rSD) and virtual-photon (vSD) corrections and $\Gamma^{(0)}_{\tau_{P2}}$ is the rate in absence of radiative corrections, 
\begin{equation}
\Gamma^{(0)}_{\tau_{P2}}=\frac{G_F^2 |V_{uD}|^2F_P^2 }{8\pi} M_\tau^3\left(1-\frac{m_P^2}{M_\tau^2}\right)^2, 
\label{eq:indratetauLO}
\end{equation}
being $D=d,s$ for $P=\pi,K$, respectively. 

The matrix element of the real-photon correction reads
\begin{widetext}
\begin{align}
   i \mathcal{M}[\tau(p_\tau)&\to P(p)\nu_\tau(q)\gamma^*(k)]=G_FV_{uD}eF_P M_\tau \Gamma_\mu \bar{u}(q)(1+\gamma_5)\left[\frac{2p^\mu}{2p\cdot k+k^2}+\frac{2p_\tau^\mu-k\!\!\!/\gamma^\mu}{-2p_\tau\cdot k+k^2}\right]u(p_\tau)\nonumber \\
& -G_FV_{uD}e\Gamma^\nu \bigg\{ i F_V^P(W^2,k^2)\epsilon_{\mu\nu\rho\sigma} k^\rho p^\sigma
+ F_A^P(W^2,k^2)\left[(W^2+k^2-m_P^2)g_{\mu\nu}-2k_\mu p_\nu\right]  \nonumber \\ 
& \qquad \qquad \qquad \qquad -A_2^P(k^2)k^2g_{\mu\nu}+A_4^P(k^2)k^2(p+k)_\mu p_\nu
\bigg\}  \bar{u}(q)\gamma^\mu(1-\gamma_5)u(p_\tau) \, , 
\end{align}
\end{widetext}
where $W^2=(p_\tau-q)^2=(p+k)^2$ and $\Gamma_\mu=-\epsilon_\mu^*(k)$ for an on-shell photon. In the first line the structure-independent contribution is shown~\cite{DF}, whereas in the second and third lines we give the structure-dependent contributions in terms of the relevant form factors, which encapsulate the hadronization of the related QCD currents. At leading order in the chiral expansion, the form factors $A_2^P$ and $A_4^P$ are not independent and can be written in terms of a single form factor $B$ (only depending on $k^2$ and identical for $P=\pi,K$ at this order): $A_2(k^2) =-2B(k^2)$ and $A_4(k^2)=-2B(k^2)/(W^2-m_P^2)$.

\begin{figure}
\begin{center}
\includegraphics[scale=0.7]{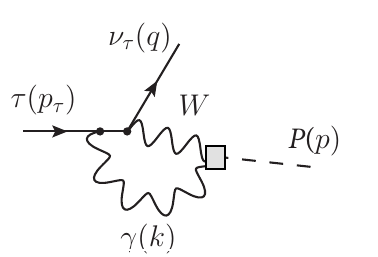} 
\end{center}
\caption{Feynman diagram corresponding to the structure dependent contributions to $\tau \to P \nu_\tau$ decays. The gray shaded box stands for the form factors. 
\label{feynman}}
\end{figure}

For virtual-photon corrections, we will focus on the structure-dependent contributions (SD), as we agree with the structure-independent ones calculated in Refs.~\cite{DF}. The relevant amplitude, where one photon vertex is attached to the $\tau$ lepton (see figure \ref{feynman}), reads 
\begin{align}
&i\mathcal{M}[\tau \to P\nu_\tau]|_{\mathrm{SD}}=%i 
G_F V_{uD}e^2 \!\int\! \frac{\textrm{d}^dk}{(2\pi)^d} \frac{\ell^{\mu\nu}}{k^2[(p_\tau\!+\!k)^2\!-\!M_\tau^2]}  \nonumber \\ & \left[i \epsilon_{\mu\nu\lambda\rho}k^\lambda p^\rho F_V^P (W^2\!,k^2)
\!+\!F_A^P(W^2\!,k^2)\lambda_{1\mu\nu}\!+\!2 B(k^2) \lambda_{2\mu\nu}\! \right]\!,\label{amp-axial-1} 
\end{align}
where we have defined
\begin{align}
\ell^{\mu\nu}&=\bar{u}(q)\gamma^\mu(1-\gamma_5)[(\cancel{p}_\tau+\cancel{k})+M_\tau]\gamma^\nu u(p_\tau),\nonumber \\
\lambda_{1\mu\nu}&=\left[(p+k)^2+k^2-m_P^2\right]g_{\mu\nu}-2k_{\mu}p_{\nu},\nonumber \\
\lambda_{2\mu\nu}&=k^2 g_{\mu\nu}-\frac{k^2 (p+k)_{\mu}p_{\nu}}{(p+k)^2-m_P^2}.
\end{align}

In our case at hand, the form factors $F^P_{V,A} (W^2, k^2)$ and $B(k^2)$ can be taken from Refs.~\cite{form_factors1,form_factors2},
\begin{eqnarray}
F^{P}_V(W^2,k^2)&=&\frac{-N_C M_V^4}{24\pi^2F_P(k^2-M_V^2)(W^2-M_V^2)}\, ,  \nonumber \\
F_{A}^P(W^2,k^2)&=&\frac{F_P}{2} \frac{M_A^2 -2M_V^2 -k^2 }{(M_V^2-k^2)(M_A^2-W^2)}\,, \nonumber \\
B(k^2)&=& \frac{F_P}{M_V^2-k^2}\,, \label{form_factors_scenariob}
\end{eqnarray}
where well-behaved two- and three-point Green functions are imposed and we consider the chiral and $U(3)$ flavor limits. In (\ref{form_factors_scenariob}) $M_V$ and $M_A$ stand for the vector and axial-vector resonance masses, $M_V=M_\rho$, $M_A=M_{a_1}$ and $M_V=M_K*$, $M_A\sim M_{f_1}$ for the pion and kaon case, respectively. \\

{\it Calculation of $\delta R_{\tau/P}=\delta_{\tau P} - \delta_{P \mu}$.} -- Adding up the structure-independent terms obtained with both virtual and real photons (SI, the point-like approximation), we confirm the results given 
in Refs.~\cite{DF}:
\begin{equation}
\delta R_{\tau/P}\big|_{\mathrm{SI}}\!=\!\frac{\alpha}{2\pi}\!\left\{\!\frac{3}{2} \mathrm{log}\!\frac{M_\tau^2 m_P^2}{m_\mu^4}\!+\!\frac{3}{2}  \!+\!g\!\left(\!\frac{m_P^2}{M_\tau^2}\!\right)\!-\!f\bigg(\!\frac{m_\mu^2}{m_P^2}\!\bigg)\!\right\},
\end{equation}
where $f(x)$ and $g(x)$ read\footnote{$F(x)$ and $G(x)$ of (\ref{eq:indrate}) and (\ref{eq:indratetau}) can be related to $f(x)$ and $g(x)$ by $F(x)=3/2 \log x + f(x)/2 +13/8 -\pi^2/3$ and $G(x)=g(x)/2+19/8-\pi^2/3$.}:
\begin{eqnarray}
f(x)&=&2\left(\frac{1+x}{1-x}\log x-2\right)\log(1-x)-\frac{x(8-5x)}{2(1-x)^2}\log x\nonumber\\
&&+4\frac{1+x}{1-x}\mathrm{Li}_2(x)-\frac{x}{1-x}\left(\frac{3}{2}+\frac{4}{3}\pi^2\right)\,,
\end{eqnarray}
and
\begin{eqnarray}
g(x)&=&2\left(\frac{1+x}{1-x}\log x-2\right)\log(1-x)-\frac{x(2-5x)}{2(1-x)^2}\log x\nonumber\\
&&+4\frac{1+x}{1-x}\mathrm{Li}_2(x)+\frac{x}{1-x}\left(\frac{3}{2}-\frac{4}{3}\pi^2\right)\,,
\end{eqnarray}
being 
Li$_2(x)=-\displaystyle\int_0^x \mathrm{d}t \frac{\log(1-t)}{t}\,.$ One gets $\delta R_{\tau/\pi}\big|_{\mathrm{SI}}= 1.05\%$ and $\delta R_{\tau/K}\big|_{\mathrm{SI}}=1.67\%$~\cite{DF}.

The  structure-dependent contributions with real photons (rSD) can be extracted from (7.14) and (7.16) of Ref.~\cite{CR2} for $P \to\mu \nu_\mu \gamma$ ($\delta_{\pi \mu}\big|_{\mathrm{rSD}}=-1.3\cdot 10^{-8}$ and $\delta_{K\mu}\big|_{\mathrm{rSD}}=-1.7\cdot10^{-5}$) and from Ref.~\cite{form_factors1} for $\tau \to P \nu_\tau \gamma$ ($\delta_{\tau \pi}\big|_{\mathrm{rSD}}=0.15\%$ and $\delta_{\tau K}\big|_{\mathrm{rSD}}=(0.18\pm 0.05 )\%$).This gives $\delta R_{\tau/ \pi}\big|_{\mathrm{rSD}}=0.15\%$ and $\delta R_{\tau/K}\big|_{\mathrm{rSD}}=(0.18\pm 0.05)\%$, being the terms from the $P$ decay negligible.

\begin{table}[t!!!!]
\begin{center}
\begin{tabular}{|c|c|c|c|}
\hline
  Contribution & $\delta R_{\tau/\pi}$   & $\delta R_{\tau/K}$ &  Ref.  \\[5pt]
\hline \hline 
SI &  $+1.05\%$& $+1.67\%$ &\cite{DF} \\
rSD  &$+0.15\%$   &$+(0.18\pm 0.05)\%$ & \cite{CR1,form_factors1} \\
vSD & $-(1.02\pm 0.57 )\%$& $-(0.88\pm 0.58)\%$ & new \\
\hline \hline
Total & $+(0.18\pm 0.57 )\%$ & $+(0.97\pm 0.58 )\%$ & new
\\
\hline
\end{tabular}
\end{center}
\caption{Numerical values of the different photonic contributions to $\delta R_{\tau/P}$: Structure Independent (SI), real Structure Dependent (rSD) and virtual Structure Dependent (vSD). Errors are not reported for contributions where the uncertainties are negligible for the level of accuracy of this analysis, that is, lower than $0.01\%$.}%}
\label{tab:tab2}
\end{table}

In the case of $P_{\mu 2}$ the structure-dependent contribution with virtual photons (vSD) can be extracted directly from (\ref{eq:indrate}) and the numerical values for $c_n^{(P)}$ of Table~\ref{tab:tab1}: $\delta_{\pi \mu}\big|_{\mathrm{vSD}}=(0.54\pm 0.12)\%$ and $\delta_{K\mu}\big|_{\mathrm{vSD}}=(0.43\pm 0.12)\%$. The new calculation we need to perform from scratch for the analysis of $R_{\tau/P}$ is the structure-dependent part with virtual photons for $\tau \to P \nu_\tau$, corresponding to the Feynman diagram of figure~\ref{feynman}. Inserting the form factors of (\ref{form_factors_scenariob}) into (\ref{amp-axial-1}) --a tedious calculation whose technical details will be explained deeper in a forthcoming longer article~\cite{future}-- yields our results
\begin{equation} \label{vSD_tau}
\delta_{\tau \pi}\big|_{\mathrm{vSD}}=-(0.48\pm 0.56) \%
,\,\delta_{\tau K}\big|_{\mathrm{vSD}}=-(0.45\pm 0.57)\%.  
\end{equation}
Accordingly, we find
\begin{equation}
\delta R_{\tau/ \pi}\big|_{\mathrm{vSD}}\!\!=\!\!-(1.02\pm 0.57 )\%, \delta R_{\tau/K}\big|_{\mathrm{vSD}}\!\!=\!\!-(0.88\pm 0.58)\%.
\end{equation}

A reliable estimation of the uncertainties in (\ref{vSD_tau}) is fundamental, since it is the most important source of error in $\delta R_{\tau/P}$. Keep in mind the great difference with the $P_{\mu 2}$ decays: (a) in $P$ decays the calculation is performed within Chiral Perturbation Theory (ChPT), so the unknown local counterterms can be determined by matching ChPT with the effective approach at higher energies, the large-$N_C$ extension including the first resonances we have quoted previously; (b) in $\tau$ decays, and due to the energy scale at hand, the calculation is done directly with the large-$N_C$ extension of ChPT, so the matching procedure to estimate the unknown counterterms is not possible anymore. Bearing in mind this handicap, we have estimated the uncertainties of $\delta_{\tau P}\big|_{\mathrm{vSD}}$ by considering two ingredients. First of all, and in order to assess the model-dependence of the effective approach, we have also calculated $\delta_{\tau P}\big|_{\mathrm{vSD}}$ with a less general scenario where only well-behaved two-point Green functions and a reduced resonance Lagrangian is used; consequently, the form factors of (\ref{form_factors_scenariob}) are different~\cite{form_factors1,form_factors2} and we take as a first source of error in (\ref{vSD_tau}) one half of the deviation in $\delta_{\tau P}\big|_{\mathrm{vSD}}$ between the two scenarios, resulting in $\pm 0.22\%$ for the pion and $\pm 0.24\%$ for the kaon. Secondly, and in order to estimate the unknown local counterterms in $\delta_{\tau P}\big|_{\mathrm{vSD}}$, whose dependences on the renormalization scale are known from our calculation, we have considered as the second source of uncertainty in (\ref{vSD_tau}) one half of the running of the counterterms between $0.5\,$and $1.0\,$GeV, giving $\pm 0.52\%$\footnote{We follow a conservative estimate of the local counterterms in (\ref{vSD_tau}), as we justify next. Seeing that the first resonances are included in the theoretical framework for $\tau$ decays, their counterterms are expected to be smaller than in $P_{\mu 2}$. However, with the effect of the running we consider here, the counterterms in the $P_{\mu2}$ case affecting $\delta_{P \mu}\big|_{\mathrm{vSD}}$ imply similar corrections to the estimation we consider in $\delta_{\tau P}\big|_{\mathrm{vSD}}$. This can be seen as a check, a posteriori, that  further running of the counterterms is not physically motivated.}. 
Adding quadratically these two uncertainties yields the errors of (\ref{vSD_tau}): $\pm 0.56\%$ and $\pm 0.57\%$ for the pion and the kaon case, respectively. \\

{\it Results} -- In Table~\ref{tab:tab2} the different contributions to $\delta R_{\tau/P}$ are summarized, leading to our final result:
\begin{equation} \label{finalresult}
\delta R_{\tau/\pi} = (0.18\pm 0.57 )\% ,  \,\,\, \delta R_{\tau/K}= (0.97\pm 0.58 )\%,
\end{equation} 
with dominant uncertainties coming from $\delta_{\tau P}\big|_{\mathrm{vSD}}$. These results should be compared with the previous ones of Refs.~\cite{DF}, $\delta R_{\tau/\pi}=(0.16\pm0.14)\%$ and $\delta R_{\tau/K}=(0.90\pm0.22)\%$. Although their central values agree remarkably, this is merely a coincidence, as the one-sigma confidence intervals agree only at the $25(38)\%$ level for the $\pi(K)$ case. In our understanding uncertainties were underestimated in Refs.~\cite{DF}, since they have approximately the size which would be expected in a purely Chiral Perturbation Theory computation. Besides, it is important to stress again that the hadronization of the QCD currents used in that work differs for real- and virtual-photon corrections, does not satisfy the high-energy behavior dictated by QCD, violates unitarity,  analyticity and the chiral limit, and a cutoff is used to regulate the loop integrals, splitting unphysically long- and short-distance regimes.

Our results can also be used to compute the radiative corrections of the individual $\tau \to P \nu_\tau[\gamma]$ decays, $\Gamma_{\tau_{P2[\gamma]}} =  \Gamma^{(0)}_{\tau_{P2}} S_{\rm ew}\left( 1 +\delta_{\tau P} \right)$, where $ \Gamma^{(0)}_{\tau_{P2}}$ was defined in (\ref{eq:indratetauLO}), $S_{\rm ew}=1.0232$ denotes the resummed universal short-distance electroweak corrections \cite{Marciano:1993sh} and $\delta_{\tau P}$ includes all remaining radiative SI and SD corrections. Considering (\ref{eq:indratetau}), $\delta_{\tau P}$ is given by
\begin{align}\label{deltaradiativetau}
&\delta_{\tau P} =  \frac{\alpha}{2\pi} \left( \! g\!\left(\!\frac{m_P^2}{M_\tau^2}\!\right) \!+\! \frac{19}{4}\! -\!\frac{2\pi^2}{3} \!-\! 3\log \frac{m_\rho}{M_\tau} \right)  \nonumber \\
&\qquad \quad \phantom{\frac{1}{2}} +  \delta_{\tau P}\big|_{\mathrm{rSD}} +  \delta_{\tau P}\big|_{\mathrm{vSD}}\,.
\end{align}
From our results one finds $\delta_{\tau \pi} = ( -0.24 \pm 0.56 ) \%$ and $\delta_{\tau K} = (-0.15 \pm 0.57) \%$.

Considering our results and the current experimental information~\cite{PDG} in (\ref{eq:MainDef}), it is found 
\begin{align} 
&\left|\frac{g_\tau}{g_\mu}\right|_\pi\!=\!0.9964 \!\pm\! 0.0028_{\mathrm{th}}\!\pm\! 0.0025_{\mathrm{exp}} \!=\! 0.9964\pm 0.0038, \nonumber \\
 &\left|\frac{g_\tau}{g_\mu}\right|_K\!=\!0.9857\!\pm\!  0.0028_{\mathrm{th}}\!\pm\! 0.0072_{\mathrm{exp}} \!=0.9857\pm 0.0078, 
\end{align}
compatible with and at $1.8\sigma$ of lepton universality, respectively. 

An interesting application is the unitarity test (see e. g. \cite{Seng:2021nar} and references therein) from the ratio
\begin{equation}\label{eq:Vusdet}
\frac{\Gamma(\tau\to K\nu_\tau[\gamma])}{\Gamma(\tau\to \pi\nu_\tau[\gamma])}=\frac{|V_{us}|^2F_K^2}{|V_{ud}|^2F_\pi^2}\frac{(1\!-\!m_{K}^2/M_\tau^2)^2}{(1\!-\!m_{\pi}^2/M_\tau^2)^2}\left(1\!+\!\delta\right),
\end{equation}
where, as a result of our calculation,  \begin{eqnarray} \label{delta_VusVud}
\delta&=&\frac{\alpha}{2\pi} \left\{\! g\!\left(\!\frac{m_K^2}{M_\tau^2}\!\right)-g\!\left(\!\frac{m_\pi^2}{M_\tau^2}\!\right) \!\right\}+\delta_{\tau K}\big|_{\mathrm{rSD}}-\delta_{\tau \pi}\big|_{\mathrm{rSD}}\nonumber \\
&&\phantom{\frac{1}{2}} +\delta_{\tau K}\big|_{\mathrm{vSD}}-\delta_{\tau \pi}\big|_{\mathrm{vSD}}= + (0.10\pm0.80)\%\,.
\end{eqnarray} 
Using the FLAG 2+1+1 result for the meson decay constants ratio $F_K/F_\pi=1.1932\pm 0.0019$~\cite{Aoki:2019cca} and masses and branching ratios from the PDG~\cite{PDG}, one gets 
\begin{equation}  \label{ourVusVud}
 \bigg|\frac{V_{us}}{V_{ud}}\bigg|\!=\!0.2288 \!\pm\! 0.0010_{\mathrm{th}} \!\pm\! 0.0017_{\mathrm{exp}} \!=\! 0.2288\pm0.0020,
\end{equation}
which is $2.1\sigma$ away from unitarity\footnote{Again we take a conservative attitude in the estimate of the uncertainties of (\ref{delta_VusVud}), since we have directly propagated those of (\ref{vSD_tau}). Alternatively, by recalculating directly the uncertainties of the difference $\delta_{\tau K}\big|_{\mathrm{vSD}}-\delta_{\tau \pi}\big|_{\mathrm{vSD}}$, that of (\ref{delta_VusVud}) drops to $\pm 0.05\%$, which would imply $\pm 0.0004_{\mathrm{th}}$ and $\pm 0.0018$ in (\ref{ourVusVud}). Taking into account that the experimental error dominates, the change is negligible and $|V_{us}/V_{ud}|$ moves from $2.1\sigma$ to $2.2\sigma$ away from unitarity, an absolutely insignificant shift.
 }, according to $|V_{ud}|=0.97373\pm0.00031$ \cite{Hardy:2020qwl}.

Alternatively, we can extract $|V_{us}|$ directly from the $\tau \to K \nu_\tau[\gamma]$ decays, $\Gamma_{\tau_{K2[\gamma]}} =  \Gamma^{(0)}_{\tau_{K2}} S_{\rm ew}\left( 1 +\delta_{\tau K} \right)$. Using our value of $\delta_{\tau K}$ given after (\ref{deltaradiativetau}), $\sqrt{2} F_K=(155.7\pm0.3)~$MeV~\cite{Aoki:2019cca}, $S_{\rm ew}=1.0232$ \cite{Marciano:1993sh}
 and masses and branching ratios from the PDG~\cite{PDG}, this yields
\begin{equation}
 |V_{us}|\!=\!0.2220 \!\pm\! 0.0008_{\mathrm{th}} \!\pm\! 0.0016_{\mathrm{exp}} \!=\! 0.2220\pm 0.0018, 
\end{equation}
at $2.6\sigma$
 from unitarity, according to $|V_{ud}|=0.97373\pm0.00031$ \cite{Hardy:2020qwl}.

We can also bind the effective couplings characterizing non-standard interactions from the $\tau \to P \nu_\tau[\gamma]$ decays, 
\begin{equation}
\Gamma(\tau\to P\nu_\tau[\gamma])=  \Gamma^{(0)}_{\tau_{P2}}  \left|\frac{\widetilde{V}_{uD}}{V_{uD}}\right|^2\!\! S_{\rm ew}\! \left( 1 +\delta_{\tau P}  + 2 \Delta^{\tau P}\right) \,,
\end{equation}
being $D=d,s$ for $P=\pi,K$, respectively. $\Delta^{\tau P}$ captures the new physics corrections and is given by\footnote{$\Delta^{\tau P}$ contains the tree-level new physics corrections that are not absorbed in $\widetilde{V}_{uD}=(1 + \epsilon^e_L + \epsilon^e_R )V_{uD}$, directly incorporated by taking 
$V_{uD}$ from nuclear $\beta$ decays~\cite{EFTtaudecays}.}~\cite{EFTtaudecays}:
\begin{equation}
\Delta^{\tau P} = \epsilon^\tau_L-\epsilon^e_L-\epsilon^\tau_R-\epsilon^e_R-\frac{m_P^2}{M_\tau(m_u+m_D)}\epsilon^\tau_P \,.
\end{equation}
Using $|V_{ud}|=0.97373\pm0.00031$ \cite{Hardy:2020qwl}, masses and branching ratios from the PDG~\cite{PDG}, $\sqrt{2} F_\pi=(130.2\pm0.8)~$MeV and $\sqrt{2} F_K=(155.7\pm0.3)~$MeV  from Ref.~\cite{Aoki:2019cca}, $S_{\rm ew}=1.0232$ \cite{Marciano:1993sh} and our values of $|V_{us}/V_{ud}|$ in (\ref{ourVusVud}) and $\delta_{\tau P}$ after (\ref{deltaradiativetau}), we find \begin{equation}
\Delta^{\tau \pi} = -(0.15\pm0.72)\cdot10^{-2}, \quad
\Delta^{\tau K} =-(0.36\pm1.18)\cdot10^{-2}
,
\end{equation}
which update the results in Refs. \cite{EFTtaudecays} for $u\leftrightarrow d$ and $u\leftrightarrow s$ transitions, respectively. These values are reported 
in the $\overline{\mathrm{MS}}$-scheme and at a scale of $\mu=2$ GeV.\\

In conclusion, our final result for $\delta R_{\tau/P}$ is consistent with the previous literature~\cite{DF}, but with much more robust assumptions, yielding a reliable uncertainty. Extracted ratios of lepton couplings are compatible with lepton universality (pion case) and  at $1.8\sigma$ (kaon case) and can also be used for testing CKM unitarity and binding effective non-standard interactions, as we have illustrated.\\

We wish to thank V. Cirigliano, M. Gonz\'alez-Alonso and A. Pich for their helpful comments and for reading the manuscript. This work has been supported in part by the Spanish Government and ERDF funds from the European Commission [FPA2017-84445-P]; by the Generalitat Valenciana [PROMETEO/2017/053]; and by the Universidad Cardenal Herrera-CEU [INDI20/13]. M.A.A.U. is funded by Conacyt through the `Estancias posdoctorales nacionales' program. G.L.C. was supported by Ciencia de Frontera Conacyt project No. 428218. G.H.T. and P.R. acknowledge the support of C\'atedras Marcos Moshinsky (Fundaci\'on Marcos Moshinsky).

%%%%%%%%%%%%%%%%%%%%%%%%%%% REFERENCES %%%%%%%%%%%%%%%%%%%%%%%%%%%%%%%%%%%

\end{document}